\documentclass[prd,final,aps]{revtex4}
\usepackage{graphicx}

\usepackage{amsmath}
\usepackage{amssymb}

\begin{document}
\title{In an expanding universe, what doesn't expand?}
\author{Richard H.~Price} 
\affiliation{Department of Physics \& Astronomy,  Center for 
Gravitational Wave Astronomy, and Center for Advanced Radio Astronomy, University of Texas at Brownsville,
Brownsville, TX 78520}
\author{Joseph D.~Romano} 
\affiliation{Department of Physics \& Astronomy and Center for 
Gravitational Wave Astronomy, University of Texas at Brownsville,
Brownsville, TX 78520}

\begin{abstract}
The expansion of the universe is often viewed as a uniform stretching
of space that would affect compact objects, atoms and stars, as well
as the separation of galaxies. One usually hears that bound systems do
not take part in the general expansion, but a much more subtle
question is whether bound systems expand partially. In this paper, a 
definitive answer is given for a very simple system: a classical
``atom'' bound by electrical attraction.  With a  mathemical
description appropriate for undergraduate physics majors, we show
that this bound system either completely follows the cosmological
expansion, or --- after initial transients --- completely ignores it.
This ``all or nothing'' behavior can be understood 
with techniques of 
junior-level mechanics. Lastly, the simple description is shown to be
a justifiable approximation of the relativistically correct
formulation of the problem.
\end{abstract}
\maketitle


\section{Introduction}\label{sec:intro} 

\smallskip
It is not hard to explain to students that the galaxies are moving
apart like pennies glued to the surface of an expanding balloon or
raisins in an expanding loaf of raisin
bread\cite{fredrickbaker,goldsmith,riordan+schramm,pricegrover,
MTWpennies}. The expanding material represents the uniform stretching
of space. But if space itself is stretching, does this mean that
everything in it is stretching? Are galaxies growing larger? Are
atoms? The usual answer is that ``bound'' systems do not take part in
the cosmological expansion.  But if space is stretching how can these
systems not be at least slightly affected? And what would it mean for
a bound system to be ``slightly affected''? Would the bound system,
for example, expand at a slower rate? While the universe expands by 
a factor of $10^6$, would a galaxy expand by, say, a factor of $10^3$?
Would less bound systems expand
more nearly with the full cosmological rate?

It turns out that these questions get a spectrum of different answers from
experts caught unprepared. Part of this confusion is the indeterminacy
of just what the question means. (See \cite{NandP} for the effects of cosmological
expansion on clusters of galaxies; see \cite{Anderson} and \cite{Cooperstocketal}
for more mathematical analyses, and for further recent references.)
In this article, we will put aside 
some subtleties, we will focus on a  clear simple question, and we will find
a clear and interesting answer.

The simple question will take the form of a simple model: a classical
``atom,'' with a negative charge of negligible mass (the ``electron'')
going around a much more massive oppositely charged ``nucleus.''  The
Coulomb binding of the atom is physically no different from the
gravitational binding of a ``small'' configuration, like a solar
system or a galaxy, but allows certain technical
simplifications\cite{simplifications}.  We will put this classical
atom in a homogeneous universe in which expansion is described by an
expansion factor $a(t)$, where $t$ is time. Our goal will be to find
the extent to which the growth of $a(t)$ causes the atom to grow,
i.e.\,, causes the electron orbit to increase in radius.

In the description of the atom, it will be useful to use two sets of
spatial coordinates,
both of them spherical polar coordinates with the
massive nucleus at the origin. The first system consists of  the ``physical'' coordinates
$r,\theta,\phi$ in which $r$ is the proper distance from the nucleus to
the electron at a given moment of time. The second set is
``cosmological'' coordinates $R, \theta,\phi$; a point at  fixed values of
$R, \theta,\phi$ is a point fixed in the stretching space of the
universe and taking part in the cosmological expansion. 
The two
coordinate systems are related by
\begin{equation}\label{rbardef} 
r=a(t)R\,.
\end{equation}
The angular
coordinates $\theta$ and $\phi $ are the same in both the physical and
the cosmological coordinates since we can think of the cosmological
expansion as proceeding radially outward from the (arbitrarily chosen) origin.

The question whether the atom takes part in the cosmological expansion
is then the question: Does the electron follow a trajectory of
bounded $r$ (no atomic expansion), or of constant $R$ (full
cosmological expansion of the atom), or does the electron do something
``in between''?

The nature of the expansion is encoded in the functional form of
$a(t)$, and the choice of this function is the choice of the
kinematics of the expanding universe. The question of what does or
does not expand is a kinematical question fundamentally unrelated to
the physics that constrains the form of $a(t)$.  For that reason we
shall use choices of $a(t) $ that lead to the clearest insights, but
we will comment on the relationship of these convenient examples to
realistic expansion laws.

The answers given by our model contain both expected and unexpected
features. An expected feature is that the comparative strengths of the
expansion and of the electrical binding determine whether the atom
expands. An unexpected feature is the ``all or nothing'' effect of
expansion. We shall see that a sufficiently loosely bound electron will
expand with the universe; it will move with constant $R$.  A more
tightly bound electron will, after some initial disturbance of its orbit,
ignore the continuing expansion and maintain bounded $r$; 
there is no intermediate behavior. We shall also see that this ``all
or nothing'' behavior makes physical sense.

This paper analyzes the expanding atom at two different levels.
First, in Sec.~\ref{sec:newt} the description uses only Newtonian
mechanics and basic electrostatics, and should be accessible to
physics students in the junior year.  Expansion effects are introduced
in this model through a very plausible heuristic stretching force in
the relatively simple differential equation for the orbital radius
$r(t)$.  This model leads to particularly clear graphical insights in
the case of cosmological expansion that is exponential in
time. Numerical results for this and another model expansion are given
to reinforce the ``all or nothing'' feature of the atomic expansion.

Second, in Sec.~\ref{sec:einstein}, the same classical atom is
analyzed using the kinematics of general relativity and Maxwell
electrodynamics in curved spacetime\cite{Bonnor}. The result of this
analysis is a differential equation for $r(t)$ that differs slightly
from that in Sec.~\ref{sec:newt}. We show, however, that the
difference is not significant. If the atom is chosen to be initially
nonrelativistic, then subsequent relativistic effects are unimportant.
Section~\ref{sec:conc} summarizes the conclusions of the paper.

\section{Newtonian analysis}\label{sec:newt}

Our model consists of an unmoving massive nucleus fixed at the origin of a spherical
polar coordinate system $r,\theta,\phi$.  The position of an electron
of mass $m$ orbiting in the equatorial plane $\theta=\pi/2 $ is
described by the functions $r(t)$, $\phi(t)$. Since only radial forces
act on the electron, its angular momentum $mr^2d\phi/dt$ is conserved,
and we define the constant of motion
\begin{equation}\label{Ldef} 
L\equiv r^2\,\frac{d\phi}{dt}
\end{equation}
to be the electron angular momentum per unit mass. In the absense of cosmological
expansion effects, the equation of motion for $r(t)
$ is 
derived in the usual way, and  takes the familiar form
\begin{equation}\label{newteqnoexp} 
\frac{d^2r }{dt^2}-\frac{L^2 }{r^3 }=-\frac{C}{r^2 }\,.
\end{equation}
The constant of electrostatic attraction 
$C$, in SI units, is $Qq/(4\pi\epsilon_0 m)$, where $Qq
$ is the magnitude of the product of the nuclear and electron 
charges.

We need now to consider introducing the effect of expansion.
According to Eq.~(\ref{rbardef}), a point fixed in the cosmological
expansion, i.e.\,, a point of constant $R,\theta,\phi$, has a radial
acceleration\cite{alsoCooperstock}
\begin{equation}\label{expacc}
\left.\frac{d^2r
}{dt^2}\right|_{\rm expansion}
=r\,\frac{d^2a/dt^2}{a}\,.
\end{equation}
It seems plausible, therefore, that we can 
treat this  term as a radial force
per unit mass, and add it to Eq.~(\ref{newteqnoexp}) to arrive 
at 
\begin{equation}\label{newteq} 
\frac{d^2r }{dt^2}-\frac{L^2 }{r^3 }=-\frac{C}{r^2 }+r\,\frac{d^2a/dt^2}{a}\,. 
\end{equation}
From the solution of this equation, and the chosen expansion factor $a(t)$, 
we can find the radial position 
$R(t)$ of the electron by using Eq.~(\ref{rbardef}). If we combine  $r(t)$ or $R(t)$ with 
 $\phi(t)$ from the integration of Eq.~(\ref{Ldef}), we arrive at a complete description
of the orbit in either physical or cosmological coordinates.

At the outset we should notice that the
comparative strength of the electrostatic and cosmological terms in
Eq.~(\ref{newteq}) can be usefully cast as a comparison of timescales
for atomic and expansion effects. We define the atomic timescale
$T_{\rm atom}$ as a combination of the parameters $L$ and $C$ relevant
to the electron's motion
\begin{equation}
  T_{\rm atom}=L^3/C^2\,,
\end{equation}
and we note that the time for the electron to complete a circular orbit, 
in the absence of expansion effects, is $2\pi T_{\rm atom}$.

We will first choose the cosmological expansion kinematics to be exponential
\begin{equation}\label{deSitter}
  a(t)=e^{t/T_{\rm exp}}\,.
\end{equation}
Such expansion, a ``de Sitter'' cosmology, is of interest in connection with inflationary models, and
mathematical relativity, but it is our first choice for a very different reason: it results in a 
form of  
Eq.~(\ref{newteq}) without any explicit time dependence:
\begin{equation}\label{newtalpha} 
\frac{d^2r }{dt^2}=\frac{L^2 }{r^3 }-\frac{C}{r^2 }+
\frac{r}{T_{\rm exp}^2}\,. 
\end{equation}
Since $t$ does not explicitly appear, 
the equation can be viewed as that for 
a particle moving in one dimension under the influence of an $r$-dependent potential.

This view is based on the fact that Eq.~(\ref{newtalpha}) guarantees 
that the energy-like quantity
\begin{equation}\label{Edef} 
E\equiv \frac{1}{2}\,\left(\frac{dr}{dt}\right)^2+\frac{L^2}{2r^2}
-\frac{C}{r}-\frac{r^2}{2T_{\rm exp}^2}
\end{equation}
is constant, so 
\begin{equation}\label{Vdef} 
V\equiv \frac{L^2}{2r^2}
-\frac{C}{r}-\frac{r^2}{2T_{\rm exp}^2}
\end{equation}
 can be viewed as an effective potential.

\begin{figure}[ht]
\includegraphics[width=.4\textwidth]{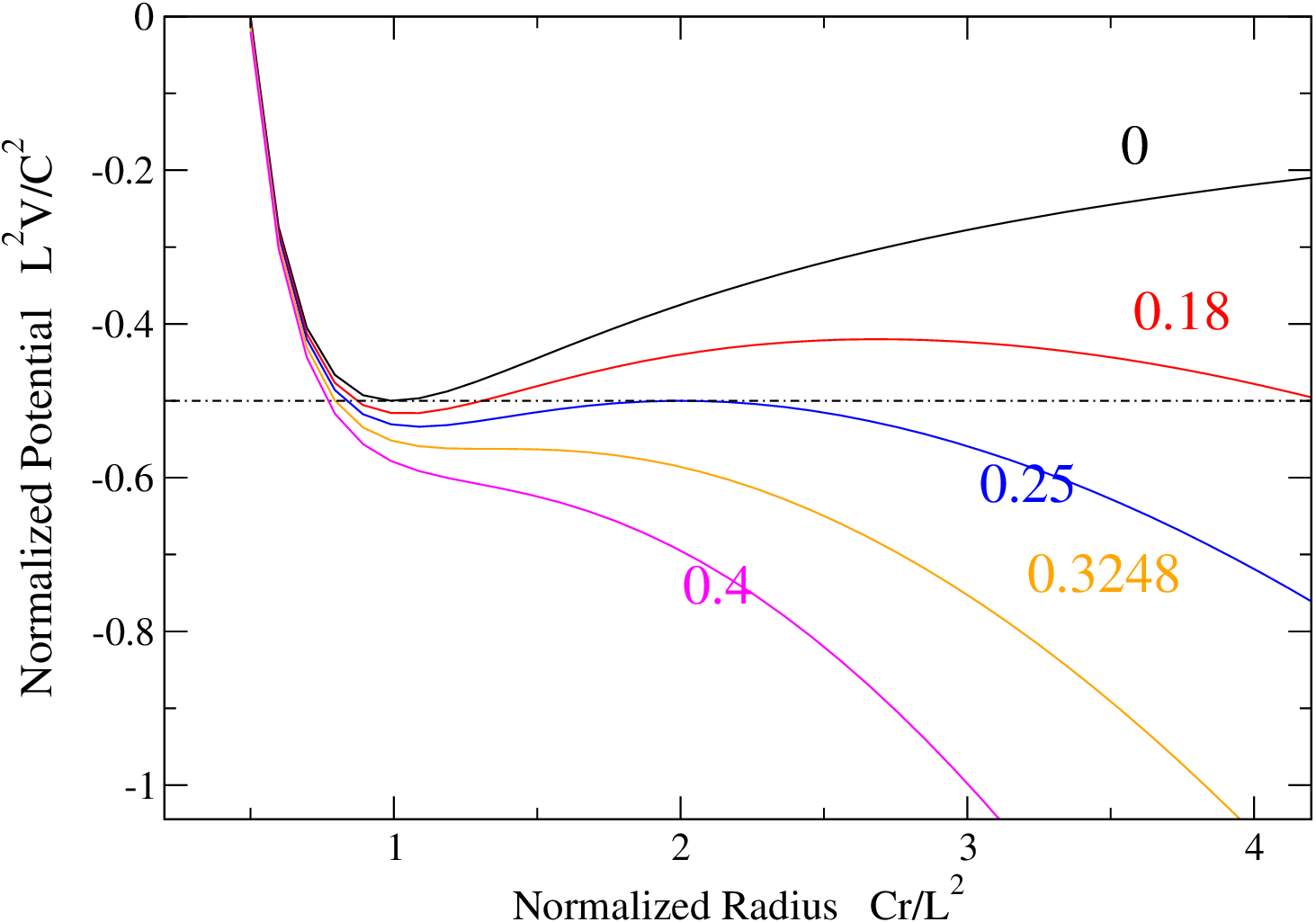} \caption{Effective
potential for exponential expansion. Curves are marked 
by the value of the parameter $T_{\rm atom}/T_{\rm exp}$. 
The curve labeled 0 is the no-expansion potential, the potential
for $T_{\rm atom}/T_{\rm exp}=0$.
The dashed line 
shows the alignment of the minimum of this no-expansion potential 
with the local maximum of the potential for 
$T_{\rm atom}/T_{\rm exp}=0.25$.
\label{fig:potentials}}
\end{figure}

Plots of this potential are given in  Fig.~\ref{fig:potentials}. Both 
the potential and the radius are made dimensionless by multiplying
them by appropriate combinations of the parameters $L$ and $C$. Each curve
is  labeled with the value of the parameter $T_{\rm atom}/T_{\rm exp}$
that determines the importance of the cosmological expansion to the 
evolution of the radius of the atom. The larger the value of 
$T_{\rm atom}/T_{\rm exp}$, the larger is the effect of expansion. 

Expansion is absent for the top curve, that for which $T_{\rm
  atom}/T_{\rm exp}=0$. In this case, the electron is always trapped in
the potential well, i.e., it is permanently bound. If it is started at
the bottom of the potential well, at $r=L^2/C$ (i.e., at
$E=-C^2/2L^2$ in Eq.~(\ref{Edef})) it will remain in a circular orbit
at that radius. For any larger value of $E$ the electron will orbit
in an
 ellipse.

For nonzero values of the $T_{\rm atom}/T_{\rm exp}$ parameter, the
potential at large $r$ eventually becomes negative and decreasing; it
represents a dominant force outward. Consequently, an electron that is
at sufficiently large radius will be driven to even larger radius. The
important question is whether the electron will ever get to that
region of dominant outward force.  The answer is contained in the
shapes of the curves in Fig.~\ref{fig:potentials}.

We first consider the case in which the electron starts at the bottom
of the potential well of the no-expansion curve, then is ``surprised''
by the sudden turn on of expansion, so that the electron has energy $E=-C^2/2L^2$
and finds itself
under the influence of one of the curves with $T_{\rm atom}/T_{\rm
  exp}>0$. For this scenario a critical
value of $T_{\rm atom}/T_{\rm exp}$ is 0.25. As shown by the dashed
line in Fig.~\ref{fig:potentials}, this is the value for which the
local peak of the potential has the same value as the lowest point in
the potential well of the no-expansion curve. For values of $T_{\rm
  atom}/T_{\rm exp}$ less than 0.25 the surprised electron will remain trapped
in an approximately elliptical orbit. For larger values of $T_{\rm
  atom}/T_{\rm exp}$ the electron will move to larger radius and be
accelerated outward by the cosmological expansion.

A somewhat different scenario can be envisioned: the electron finds itself 
in the bottom not of the well of the no-expansion curve, but at the bottom
of the potential well of the curve that includes the expansion term.
In this case the electron will remain at a fixed value of $r$, the location 
of the bottom of the potential well,  but this can 
happen only if there actually exists such a potential well. As shown in 
Fig.~\ref{fig:potentials} there is a critical curve that separates potentials
with and without a potential well. That curve turns out to correspond 
to $T_{\rm atom}/T_{\rm exp}=3\sqrt{3\;}/16\approx0.3248$.

With the viewpoint of the potentials it is clear why there is an ``all
or nothing'' behavior of the atom. The electron either is, or is not,
trapped in the potential well; there is no ``partial expansion''
possible.  Underlying this graphical understanding is a broader but
less precise heuristic explanation of the ``all or nothing'' effect,
an explanation that goes beyond expansion that is exponential in
time. The cosmological expansion term $r(d^2a/dt^2)/a$ increases at
large physical distances $r$ from the nucleus. The centrifugal and
electrical forces decrease. This implies a sort of instability with
respect to expansion. If the electron moves sufficiently far outward,
the expansion term will become more important and push the electron
yet further outward.

\begin{figure}[ht]   
\includegraphics[width=.4\textwidth]{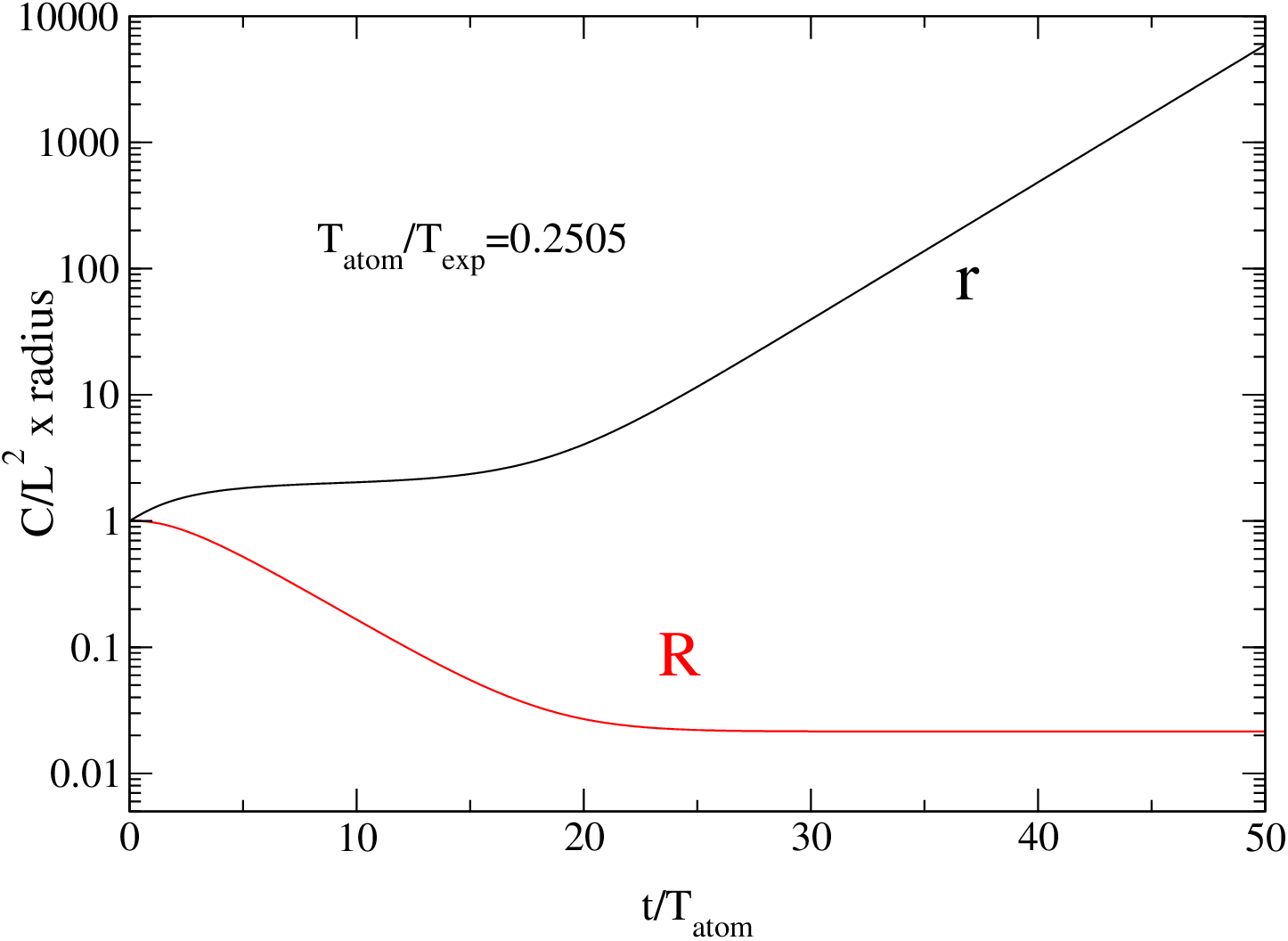}
\hspace{.05\textwidth}
\includegraphics[width=.42\textwidth]{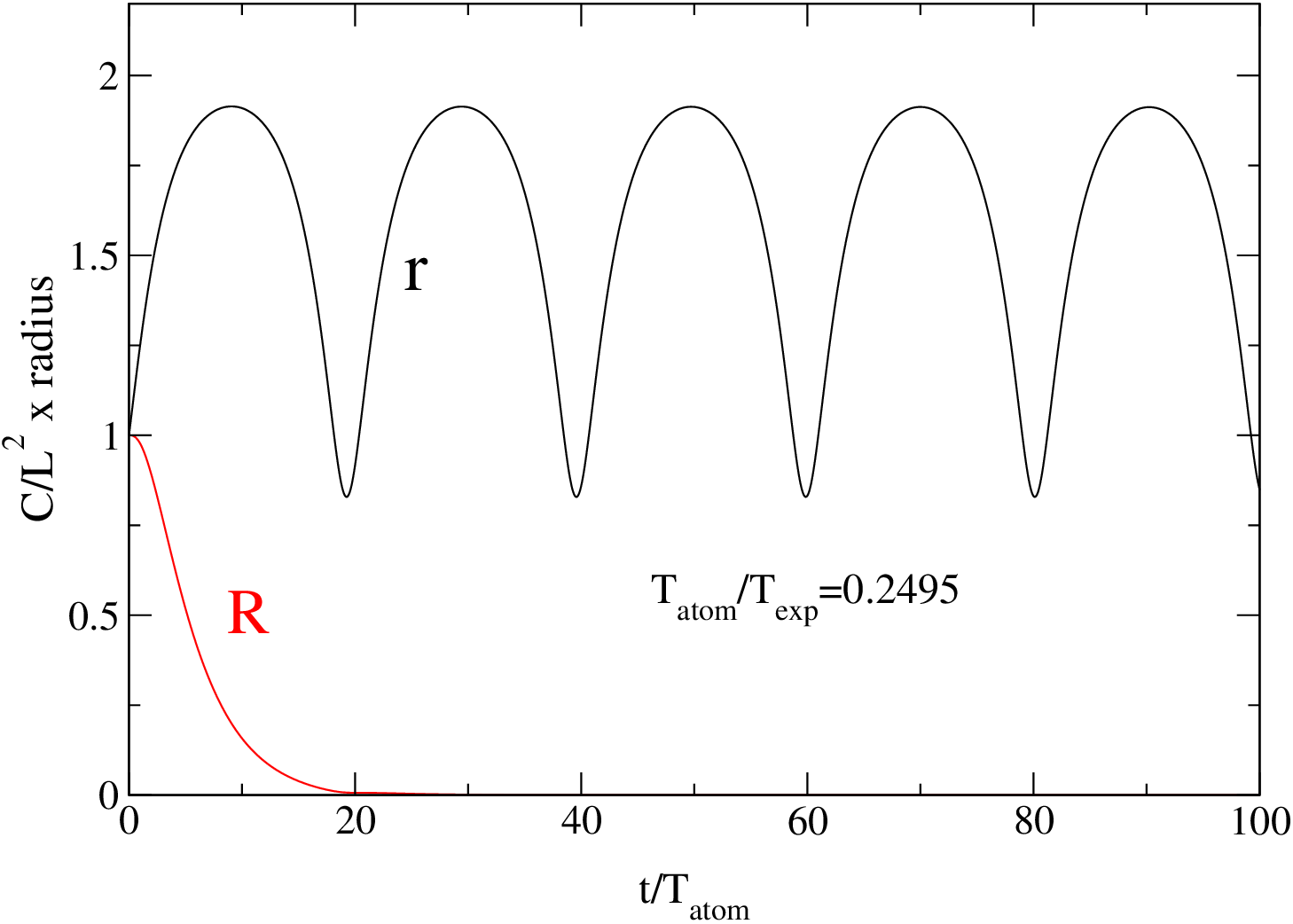}
\caption{Radius as a function of time for exponential expansion.  On
  the left is the case for $T_{\rm atom}/T_{\rm exp}=0.2505$ for which
  the electron cosmological radius $R$ remains approximately constant
  after an initial decrease to about 2\% of its initial value. Due
  to the exponential increase in $a(t)$, the physical radius $r$ grows
  without bound.  On the right is the radial kinetics for a slighly
  smaller value, 0.2495, of $T_{\rm atom}/T_{\rm exp}$. In this case the
  electron remains bound in an approximately elliptical orbit with the physical
  radius oscillating between values near the original atomic radius.
  The coordinate radius $R$ in this case falls off exponentially. 
\label{fig:expwith2Ls}}
\end{figure}

We can get another viewpoint on the bound vs.~unbound issue by solving
Eq.~(\ref{newtalpha}), the equation of motion for the electron,
computationally.  In principle we could start the computation with the
electron at the bottom of a potential well for expansion. The results
turn out to be in agreement with the predictions of the analysis based
on Fig.~\ref{fig:potentials}; the electron stays at a constant value
of $r$, so this result is not of particular interest.  More
interesting is the surprised electron scenario, the case of an
electron, in an expanding universe, with initial $r$ and $dr/dt$ such
that the electron's energy, according to Eq.~(\ref{Edef}), corresponds
to the bottom of the no-expansion potential well, i.e., the electron
energy is $-C^2/2L^2$.  The results, shown in
Figs.~\ref{fig:expwith2Ls}, are in accord with the analysis based on
Fig.~\ref{fig:potentials}. (The energy corresponds to the dashed horizontal
line in Fig.~\ref{fig:potentials}.) For $T_{\rm atom}/T_{\rm exp}$ slightly
greater than the 0.25 critical value, $r$ the physical radius of the
atom, grows exponentially after an initial hesitation.  By contrast,
for $T_{\rm atom}/T_{\rm exp}$ slightly less than this critical value
the electron remains trapped in an approximately elliptical orbit, and
is unaffected by the exponential expansion.

It is important to check that our understanding, based on exponential 
expansion, applies for other expansion laws.
This check will be based on $a(t)$ roughly proportional to $t^2$. For convenience we
will choose the actual dependence to be 
\begin{equation}\label{modifiedt2law} 
a(t)=1+ \left(\frac{t}{T_{\rm exp}}\right)^2      \tanh(t/T_{\rm exp})\,.
\end{equation}
For $t$ much larger than $T_{\rm exp}$ this expansion factor is proportional to $t^2$
but its properties at $t=0$ simplify our considerations.
Both $da/dt$ and $d^2a/dt^2$ vanish at $t=0$, so we can start 
the expansion with both $dr/dt=0$ and $dR/dt=0$. In addition, the expansion 
term in Eq.~(\ref{newteq}) vanishes at $t=0$, so there will be no initial 
cosmological acceleration;
if we choose the  condition 
 $r=L^2/C$ for balance of Coulomb and centripetal force, then the 
there will be no initial acceleration.

\begin{figure}[ht]   
\includegraphics[width=.4\textwidth]{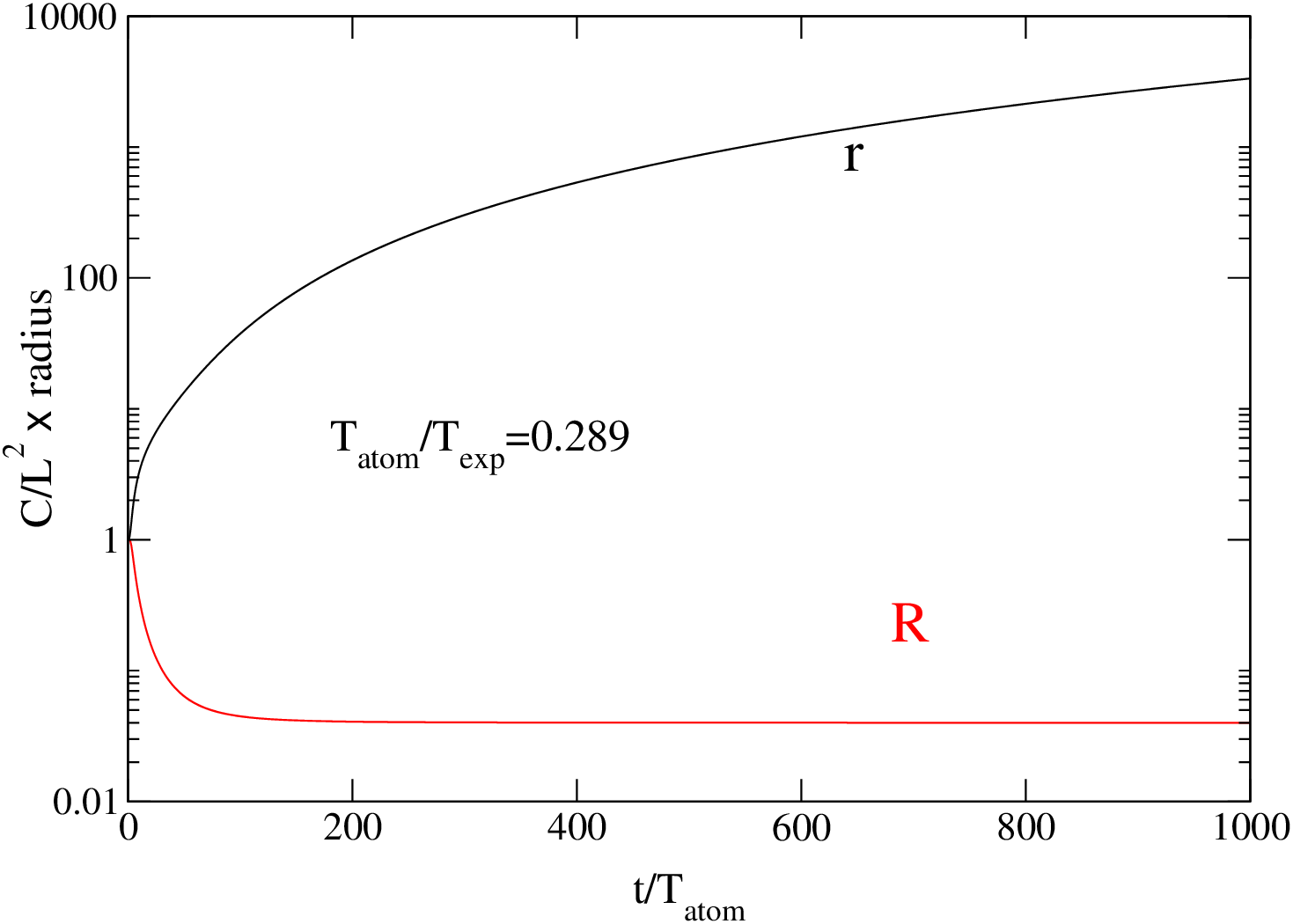}
\hspace{.05\textwidth}
\includegraphics[width=.42\textwidth]{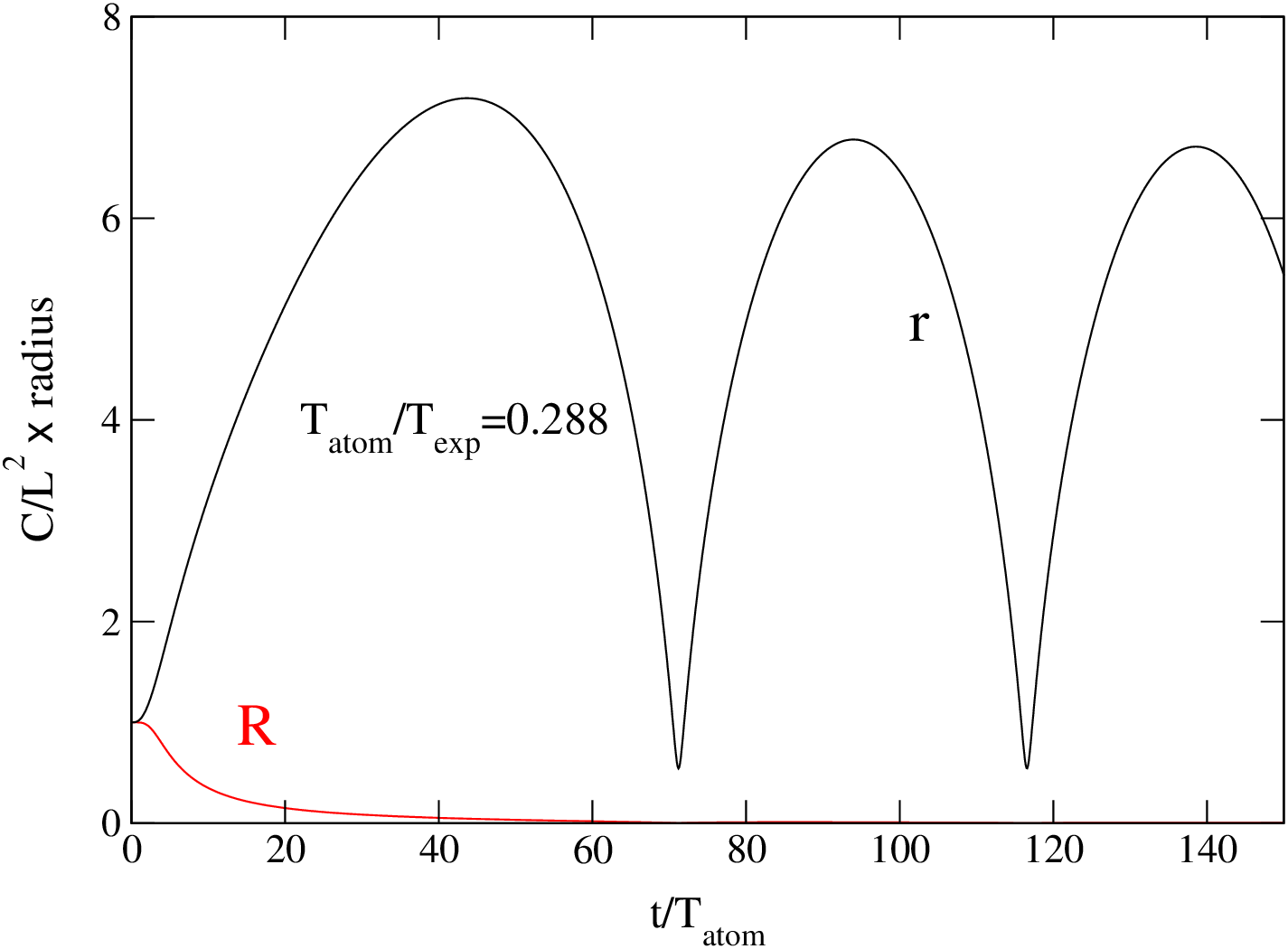}
\caption{Radius as a function of time for the modified $a\propto t^2$
  expansion described in the text. The figure on the left is for
  $T_{\rm atom}/T_{\rm exp}=0.289$ for which the electron is unbound. In
  this case the cosmological radius $R$ remains constant for large times at
  about 4\% of its initial value, while the physical radius $r$
  expands proportional to $t^2$. On the right is shown the radii as a
  function of time for $T_{\rm atom}/T_{\rm exp}=0.288$. Here the
  cosmological radius $R$ decreases asymptotically to zero roughly as
  $t^{-2}$, while the physical radius oscillates as the electron orbits in a bound, approximately
  elliptical, orbit.
\label{fig:Newtt2exp} }
\end{figure}

Results are shown in Fig.~\ref{fig:Newtt2exp} for two very nearby
values of the parameter $T_{\rm atom}/T_{\rm exp}$. 
For the expansion law of Eq.~(\ref{modifiedt2law})
we see the same qualitative phenomenon as for 
the exponential expansion: the atom either fully takes
part in the cosmological expansion or, for a slightly 
 smaller value of $T_{\rm atom}/T_{\rm exp}$, it remains bound.


\section{Relativistic analysis}\label{sec:einstein}

The analysis in the previous section was based on a heuristic
term in Eq.~(\ref{newteq}) representing the effect of expansion. 
Here we analyze the problem using relativisitic cosmology and Maxwell-Einstein
theory. 

We start with a standard form\cite{MTW} for the spacetime metric of 
a homogeneous isotropic universe
\begin{equation}\label{FRW} 
ds^2=-c^2dt^2+a^2(t)\left[\frac{dR^2}{1-kR^2}
+R^2\left(d\theta^2+\sin^2\theta d
\phi^2\right)\right]\,.
\end{equation}
Here, as in Sec.~\ref{sec:newt}, $a(t)
$ is the expansion factor and $R$ is the cosmological radius, with 
$r=a(t)R$ the physical radius. 
As we shall explain in more detail below,
the presence of the speed of light, $c$,
in the line element
introduces an additional parameter for 
relativistic motion, which, 
for our classical atom, is the ratio 
of the initial orbital speed of the electron
to the speed of light.

The constant $k$ can be positive, negative, or zero, and has a
magnitude of order $1/R_c^2$, where $R_c$ is a characteristic
cosmological distance. If $R^2 /R_c^2$ is not negligibly small, it
means that our atom occupies a significant fraction of the
universe. For our considerations of ``what expands'' we want
our atom to be very small compared to the size of the universe, so we
omit the $kR^2$ term in Eq.~(\ref{FRW}), i.e., we set $k=0$.

The first step in the relativistic analysis is to find the correct
description of the electrical attraction. For the spherically
symmetric electromagnetic field of the nucleus there can only be a
component $F^{0R}$ of the electromagnetic tensor $F^{\mu\nu}$. The Maxwell
equations $F^{\alpha\beta}_{;\beta}=0$, with $\alpha=0$ and with $\alpha=R$
give us
\begin{equation}\label{divlaw}
\frac{1}{R^2}\,\left(F^{0R}{R^2}
\right)_{,R}
=0=
\frac{1}{a^3(t)
}\,\left(F^{0R}a^3(t)
\right)_{,t}
\end{equation}
so that the solution must have the form
\begin{equation}\label{F0Req}
F^{0R}=
\frac{Q}{R^2a^3}\,.
\end{equation}

The $R$ equation of motion of the electron's 4-velocity $U^\alpha$ is
\begin{equation}\label{coulomblaw}
U^\alpha U^R_{;\alpha}=\frac{q}{m}U_0F^{0R}\,,
\end{equation}
where $q$ is the magnitude of the charge of the electron.
For motion in the $\theta=\pi/2 $ plane this becomes, after some
manipulations,
\begin{equation}\label{einsteq} 
\frac{d}{dt}\left(a^2
\frac{U^0}{c}\,\frac{dR}{dt}
\right)
-\frac{L^2}{a^2R^3(U^0/c)}
=-\frac{C}{aR^2}\,.
\end{equation}
Here $L\equiv U_\phi=r^2\, (U^0/c)\, d\phi/dt$
is a constant of the motion, and we have chosen the constant $C
$ to be analogous to the same symbol in Eq.~(\ref{newteq}).
We now note that Eq.~(\ref{newteq}), with $r=a(t)R$, can 
be written in the form
\begin{equation}\label{newtapptoee}
  \frac{d}{dt}\left(a^2
\frac{dR}{dt}
\right)
-\frac{L^2}{a^2R^3}
=-\frac{C}{aR^2}\,.
\end{equation}
The adequacy of the Newtonian analysis therefore depends on 
the extend to which ${U^0}/{c}$ differs from unity. 

\begin{figure}[ht]
\includegraphics[width=.4\textwidth]{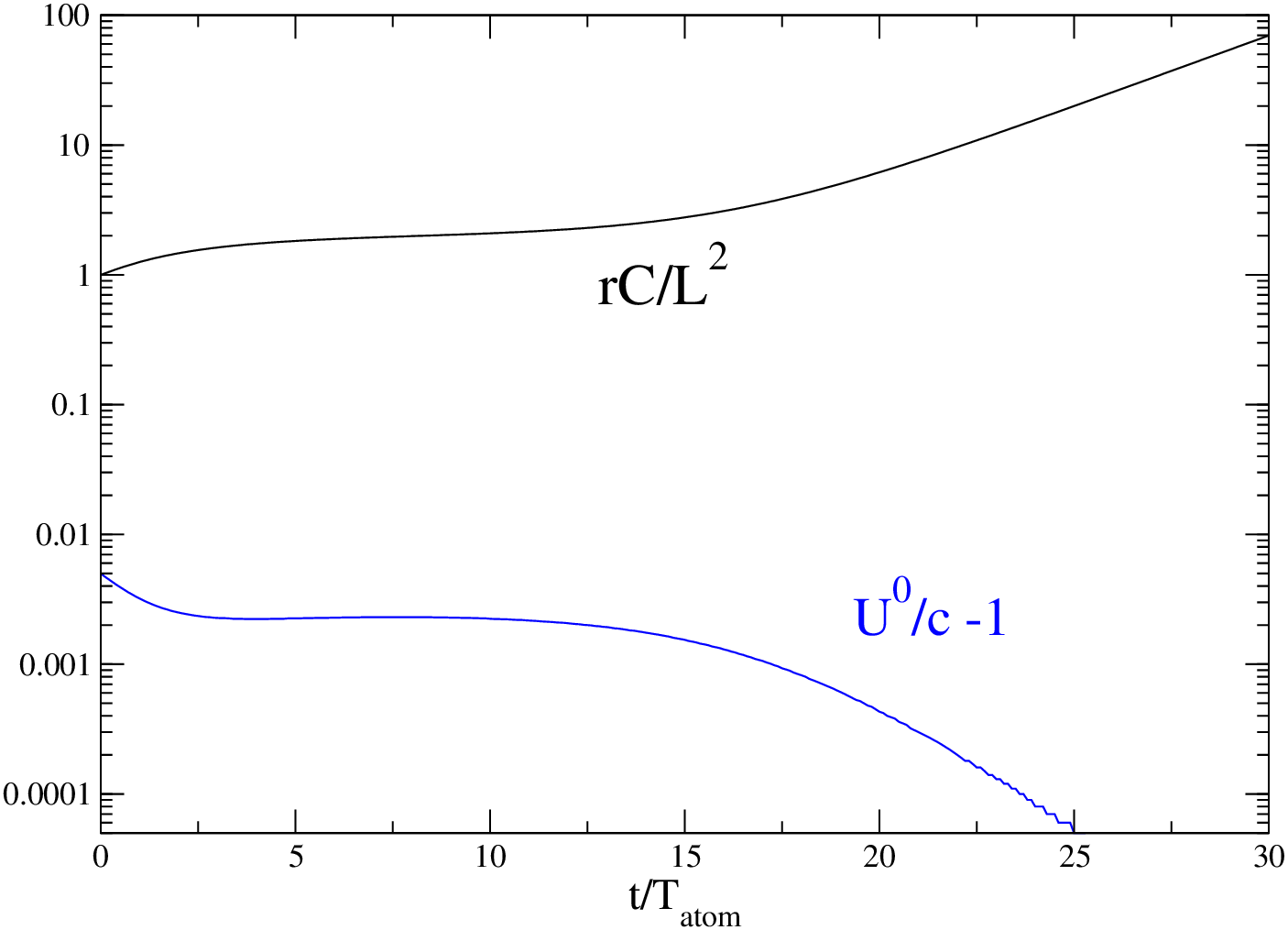} \caption{
The unbound physical radius and the index of relativistic 
effects ${U^0}/{c}-1$ for exponential expansion
with $T_{\rm atom}/T_{\rm exp}=0.252$,
starting
with $dR/dt=0$ and $\beta_0=0.1$. 
\label{fig:U0}}
\end{figure}

To compute motion for a  relativistic model,
Eq.~(\ref{einsteq}) must be solved simultaneously
with an expression for $U^0/c$.
This additional expression can be obtained from
the normalization of the 4-velocity,
$U^\mu U_\mu=-c^2$, leading to
\begin{equation}
\frac{U^0}{c}=
\left[
1 
- \left(\frac{a}{c}\frac{dR}{dt}\right)^2
- \left(\frac{aR}{c}\frac{d\phi}{dt}\right)^2
\right]^{-1/2}\,.
\label{U0explicit}
\end{equation}
An implicit expression for $U^0/c$, 
which is useful for understanding the subsequent 
time-evolution of $U^0/c$, is
\begin{equation}\label{U0eq} 
\frac{U^0}{c}=
\sqrt{1+\left(\frac{a}{c}\frac{U^0}{c}\frac{dR}{dt}\right)^2
+\left(\frac{L}{aRc}\right)^2\;}
\,.
\end{equation}

In the Newtonian case a model for the classical atom
required only the choice of the expansion law, and a value of a single
dimensionless parameter $T_{\rm atom}/T_{\rm exp}$. For relativistic
motion there is an important difference; we must now choose a second
dimensionless parameter $\beta_0=v_0/c$, where $v_0=C/L$ is the initial
orbital speed of the electron times $2\pi$. 
This need for a second parameter is
instructive. If we were to fix, say, $T_{\rm atom}=T_{\rm exp}$, this could
correspond to slow electron motion (compared to $c$) and slow expansion,
or to fast electron motion and fast expansion. In the second case, but not 
the first, relativistic effects would be important. 

If $\beta_0$ is not chosen small compared to unity, then relativistic
effects will be important even initially. Such effects, while
interesting in their own right, are not related to cosmological
expansion, and are not our focus here. Rather, what is of primary
interest is the question whether an atom that is not initially
relativistic can become relativistic when it is cosmologically
expanding. We investigate this first numerically, with a universe
following the exponenetial expansion in Eq.~(\ref{deSitter}).  We
start with $dR/dt=0$, and choose $\beta_0=0.1$ so that the electron starts out mildly
relativistic and we can follow the evolution of its relativistic
effects.  Since we want the atom to be unbounded, we take $T_{\rm
  atom}/T_{\rm exp}$ to be 0.252. (It turns out that with
$\beta_0\neq0$ the unbound behavior requires a slightly larger value
of $T_{\rm atom}/T_{\rm exp}$ than in the Newtonian case.)

Results for this model are shown in Fig.~\ref{fig:U0}. The plots show
the unbounded growth of the physical radius $r$, and show that
$U^0/c-1$, the measure of the relativistic nature of the electron
motion, {\em decreases} with the expansion of the atom. The
mathematical basis for this is not hard to see in Eq.~(\ref{U0eq}). At
large expansion, Eq.~(\ref{einsteq}) tells us that the combination
$a^2(U^0/c)dR/dt$ is approximately constant.  This means that the
middle term inside the square root of Eq.~(\ref{U0eq}) must fall off
as $a^{-2}$.  The last term in the square root also falls off with the
expanison.  The implication, validated by Fig.~\ref{fig:U0}, is that
$U^0/c-1\rightarrow 0$ with unbounded expansion.

The mathematical ``how'' is then clear, but the physical ``why'' must
be explained. In this connection it is interesting to consider the
velocity of an unbound electron relative to the ``fabric of the universe,''
i.e., the velocity $v_{\rm loc}$ that would be measured in the local
Minkowski frame of an observer comoving with the cosmological
flow\cite{BunnHogg}, an observer with constant $R, \theta$, and
$\phi$. That velocity is easily shown to be
\begin{equation}
  v_{\rm loc}=c\sqrt{1-({c}/{U^0)^2}\;}\,.
\end{equation}
As the expansion proceeds, the particle is, in some sense,
becoming less and less relativistic.


\section{Conclusion}\label{sec:conc}

We have presented a simple definitive question about the influence of
the expansion of the universe on a very particular system: a classical
``atom.''  And we have found a simple definitive answer: Expansion forces
increase with increasing atomic radius, while atomic forces decrease. 
This 
amounts to an instability with respect to the disruption of an atom. If the 
atomic accelerations are initially stronger than the cosmological, then 
the subsequent expansion will become less and less important. The atom 
will not ``partially'' take part in the expansion. If, on the other hand,
the cosmological effect is initially stronger, the atomic radius will increase
and the atomic forces will become less and less important. The atom will
fully take part in the expansion.

In analyzing this problem we have relied on a simple description of
expansion, that of Eq.~(\ref{newteq}), that avoids relativistic
effects.  A major pedagogical point is the simple graphical way in
which the ``what expands'' question can be graphically understood for
the special case of exponential cosmological expansion.

With an {\em a priori} correct general relativistic  calculation, and 
numerics, we have shown that  the simple nonrelativistic 
model is fully adequate. We have also shown that for an atom
that is not bound, but that expands with the universe, relativistic 
effects become {\em less} important as the atom gets larger.

We end with a ``practical'' consideration.
Our quantification of the relative strengths of atomic and expansion
forces was given in terms of a characteristic time $T_{\rm atom}$ for
the motion of electrons in atoms, and $T_{\rm exp}$, the cosmological
expansion time, e.g., the Hubble time. Our analyses showed that atomic
forces are initially stronger if $T_{\rm atom}/T_{\rm exp}$ is less
than order unity. Since $T_{\rm atom}$ is around $10^{-16}$ sec., and
$T_{\rm exp}$ is around $4\times10^{17}$\,sec., atoms are in no danger
of being disrupted by cosmological expansion.

\section{Acknowledgment}\label{sec:ack} This paper is the result of a
question posed by high school student Deepak Ramchand Mahbubani, Jr.
at UTB's ``21st Century Astronomy Ambassador's Program,'' and by 
the lack of a clear answer at the right level.
JDR acknowledges support from NSF grants CREST-0734800 and PHY-0855371.
We thank Prof. Bonnor for bringing to our attention Ref.~\cite{Bonnor}.



\begin{thebibliography}{}
\bibitem{simplifications} Indeed, there are no simplifications needed in the Newtonian
approximation. For the relativistic case, however, a separation of the cosmological
and local gravitational effects is necessary for a simple analysis.
\bibitem{alsoCooperstock} This heuristic Newtonian result
agrees  with Eq.~(2.8) of the general relativistic analysis in
Ref.~\cite{Cooperstocketal} for the 
acceleration in Fermi normal coordinates induced by uniform expansion.
Note that the analysis in that reference is a valid approximation 
only for distances from the nucleus small compared to a characteristic
cosmological distance. As shown in Sec.~\ref{sec:einstein}, our 
Eq.~(\ref{newteq}) applies without such a constraint.
\end{thebibliography}
\end{document}